\documentclass{elsart}
\usepackage{natbib}
\usepackage{epsfig}

\makeatletter
\def\alt{\mathrel{\mathpalette\vereq<}}
\def\vereq#1#2{\lower3pt\vbox{\baselineskip1.5pt \lineskip1.5pt
\ialign{$\m@th#1\hfill##\hfil$\crcr#2\crcr\sim\crcr}}}
\def\agt{\mathrel{\mathpalette\vereq>}}
\def\thebibliography#1{\section*{REFERENCES}\list{\arabic{enumi}.}
  {\settowidth\labelwidth{#1.}\leftmargin=2em
   \labelsep\leftmargin \advance\labelsep-\labelwidth
   \itemsep\z@ \parsep\z@
   \usecounter{enumi}}\def\makelabel##1{\rlap{##1}\hss}%
   \def\newblock{\hskip 0.11em plus 0.33em minus -0.07em}
   \sloppy \clubpenalty=4000 \widowpenalty=4000 \sfcode`\.=1000\relax}
\makeatother

\begin{document}
\runauthor{Raffelt}
\begin{frontmatter}
\title{Neutrino Masses in Astroparticle 
Physics}

\author{Georg G.~Raffelt}

\address{Max-Planck-Institut f\"ur Physik
  (Werner-Heisenberg-Institut),
  F\"ohringer Ring 6, 80805 M\"unchen, Germany}
                           
\begin{abstract}
The case for small neutrino mass differences from atmospheric and
solar neutrino oscillation experiments has become compelling, but
leaves the overall neutrino mass scale $m_\nu$ undetermined.  The most
restrictive limit of $m_\nu<0.8~{\rm eV}$ arises from the 2dF galaxy
redshift survey in conjunction with the standard theory of
cosmological structure formation. A relation between the hot dark
matter fraction and $m_\nu$ depends on the cosmic number density
$n_\nu$ of neutrinos.  If solar neutrino oscillations indeed
correspond to the favored large mixing angle MSW solution, then
big-bang nucleosynthesis gives us a restrictive limit on all neutrino
chemical potentials, removing the previous uncertainty of $n_\nu$.
Therefore, a possible future measurement of $m_\nu$ will directly
establish the cosmic neutrino mass fraction $\Omega_\nu$.
Cosmological neutrinos with sub-eV masses can play an interesting role
for producing the highest-energy cosmic rays (\hbox{$Z$-burst}
scenario).  Sub-eV masses also relate naturally to leptogenesis
scenarios of the cosmic baryon asymmetry. Unfortunately, the
time-of-flight dispersion of a galactic or local-group supernova
neutrino burst is not sensitive in the sub-eV range.
\end{abstract}


\end{frontmatter}

{\em Prepared for the Dennis Sciama Memorial Volume of 
New Astronomy Reviews (NAR), edited by F.~Melchiorri, J.~Silk,
and Y.~Rephaeli.}


\section{Introduction}

Atmospheric and solar neutrino experiments provide rather compelling
evidence for the phenomenon of flavor oscillations.  The celebrated
up-down-asymmetry of the atmospheric $\nu_\mu$ flux measured by
Super-Kamiokande is consistently explained by $\nu_\mu\to\nu_\tau$
oscillations \citep{Fukuda2000} with the mixing parameters that are
summarized in Table~\ref{tab:osci}. The K2K long-baseline experiment
provides a first laboratory confirmation, albeit in a pure
disappearance mode \citep{Nishikawa2002}. The recent results from SNO
have largely established active-active flavor oscillations as a
solution of the solar neutrino problem
\citep{Ahmad2002a,Ahmad2002b}. The LMA parameters are strongly
favored, but the LOW case may still be viable
(Table~\ref{tab:osci}). Neutrino mass differences that are small
compared to the eV scale seem to be established.

\begin{table}
\caption{Experimental evidence for neutrino flavor
oscillations.\label{tab:osci}}
\smallskip
\begin{tabular}[4]{llll}
\hline\noalign{\vskip2pt}\hline\noalign{\vskip2pt}
Evidence&Channel&$\Delta m^2$ [$\,\rm eV^2$]
&$\sin^22\Theta$\\
\noalign{\vskip2pt}\hline\noalign{\vskip2pt}
Atmospheric&$\nu_\mu\to\nu_\tau$&(1.6--3.9)${}\times10^{-3}$&0.92--1\\
Solar\\
\quad LMA
&$\nu_e\to\nu_{\mu\tau}$&
(0.2--2)${}\times10^{-4}$&0.2--0.6\\
\quad LOW&$\nu_e\to\nu_{\mu\tau}$
&$1.3\times10^{-7}$
&0.92\\
LSND&$\bar\nu_\mu\to\bar\nu_e$&0.2--10&(0.2--$3)\times10^{-2}$\\
\noalign{\vskip2pt}
\hline
\noalign{\vskip2pt}
\hline
\end{tabular}
\bigskip
\bigskip
\end{table}

The only spanner in the works of this beautiful interpretation is the
persistence of the unconfirmed evidence for flavor transformations
from the LSND experiment~\citep{Eitel2000}. If interpreted in terms of
neutrino oscillations, the mixing parameters from all three sources of
evidence are mutually inconsistent. Even including a putative sterile
neutrino no longer provides a good global fit because all three
sources of evidence prefer active-active over active-sterile
oscillations \citep{Strumia2002}. It is expected that MiniBooNE at
Fermilab will confirm or refute LSND within the upcoming two years
\citep{Tayloe2002}.

As there is no straightforward global interpretation of all
indications for neutrino oscillations I will follow the widespread
assumption that something is wrong with the LSND signature.  If it is
due to neutrino conversions after all, something fundamentally new is
going on in the neutrino sector.  In that case much of the current
thinking in this field will have to be revised.

This is certainly the attitude that Dennis Sciama would have taken.
He eloquently advocated a cosmological scenario of radiatively
decaying dark-matter neutrinos which solves several astrophysical
problems, but requires several flavors of light sterile neutrinos and
anomalously large electromagnetic transition moments
\citep{Adams1998,Sciama1998}.  No doubt he would have challenged the
more conservative approach taken here. The final verdict on neutrino
masses and mixings is not yet in, let alone on the more exotic
possibilities.  The phenomenology of the neutrino sector may turn out
to be far richer than envisaged in my presentation of a rather minimal
scenario.

In what follows I will always assume that there are three neutrino
mass eigenstates separated by the atmospheric and solar mass
differences. In this scenario, a number of obvious questions remain
open. The 12 and 23 mixing angles are large, the 13 mixing angle is
small, but how small? Are there CP-violating phases in the mixing
matrix? Are the neutrino masses of Dirac or Majorana nature?  Is the
ordering of the masses ``normal'' with $m_2^2-m_1^2$ corresponding to
the solar and $m_3^2-m_2^2$ to the atmospheric splitting, or is it
inverted?  And finally, what is the overall neutrino mass scale? Are
the masses hierarchical with $m_1\ll m_2\ll m_3\approx 50~{\rm meV}$
or degenerate with $m_1\approx m_2\approx m_3\gg 50~{\rm meV}$?

I will focus on these last questions and review the implications of
neutrino masses in astrophysics and cosmology. Traditionally,
cosmology has provided the most restrictive limits on neutrino masses,
and this is again the case using large-scale galaxy redshift surveys
in conjunction with the standard theory of structure formation.
Conversely, if the solar LMA solution is indeed correct, the cosmic
neutrino number density is well constrained by big-bang
nucleosynthesis so that a laboratory measurement of the absolute
neutrino masses, for example in the KATRIN tritium experiment
\citep{Osipowicz2001}, would directly establish the cosmic neutrino
mass fraction. Moreover, neutrino masses can have a number of other
interesting implications in astroparticle physics in the context of
cosmic-ray physics, cosmological baryogenesis, and SN physics.

In Sec.~2 this review begins with neutrino dark matter and the latest
$m_\nu$ limits from large-scale redshift surveys.  Sec.~3 turns to the
related question of how many neutrinos there are in the universe and
how this issue connects with the solar neutrino problem.  Sec.~4 deals
with $Z$-burst scenarios for producing the highest-energy cosmic rays,
Sec.~5 with leptogenesis scenarios for producing the baryon asymmetry
of the universe. Sec.~6 is devoted to the time-of-flight dispersion of
supernova neutrinos caused by a non-vanishing $m_\nu$.  Finally, Sec.~7
summarizes the status of neutrino masses in astroparticle physics.


\section{Neutrino Dark Matter and Cosmic Structure Formation}

The cosmic number density of neutrinos and anti-neutrinos per flavor
is $n_{\nu}=\frac{3}{11}\,n_\gamma$ with $n_\gamma$ the number density
of cosmic microwave photons, and assuming that there is no neutrino
chemical potential. With $T_\gamma=2.728~{\rm K}$ this translates into
$n_{\nu}=112~{\rm cm}^{-3}$. If neutrinos have masses one finds a
cosmic mass~fraction
\begin{equation}\label{eq:Oega_nu}
\Omega_{\nu}h^2=\sum_{\rm flavors}\frac{m_\nu}{92.5~{\rm eV}}\,,
\end{equation}
where as usual $h$ is the Hubble constant in units of $100~\rm
km~s^{-1}~Mpc^{-1}$.  The requirement that neutrinos do not overclose
the universe then leads to the traditional mass limit $\sum m_\nu\alt
40~{\rm eV}$, an argument that was first advanced in a classic paper
by \citet{Gershtein1966}.

Later \citet{Cowsik1973} speculated that massive neutrinos could
actually constitute the dark matter of the universe. However, it soon
became clear that neutrinos were not a good dark matter candidate for
two reasons. The first argument is based on the limited phase space
for neutrinos gravitationally bound to a galaxy \citep{Tremaine1979}.
As a consequence, if massive neutrinos are supposed to be the dark
matter in galaxies, they must obey a {\em lower\/} mass limit of some
30~eV for typical spirals, and even 100--200~eV for dwarf galaxies.

Today the most restrictive laboratory limits on the overall neutrino
mass scale arise from the Mainz \citep{Weinheimer1999} and Troitsk
\citep{Lobashev1999} tritium end-point experiments. The current
limit is \citep{Weinheimer2002}
\begin{equation}\label{eq:tritiumlimits}
m_\nu<2.2~{\rm eV}\qquad\hbox{at 95\% CL}\,.
\end{equation}
This limit applies to all mass eigenstates if we accept that the mass
differences are as small as indicated by the atmospheric and solar
oscillation interpretation.  This limit is so restrictive that
neutrinos as galactic dark matter are completely out of the question,
even without any further appeal to cosmic structure formation
arguments.

However, cosmic structure formation does place powerful limits on the
neutrino mass. The observed structure in the distribution of galaxies
is thought to arise from the gravitational instability of primordial
density fluctuations. The small masses of neutrinos imply that they
stay relativistic for a long time after their decoupling (``hot dark
matter''), allowing them to stream freely, thereby erasing the
primordial density fluctuations on small scales
\citep{Doroshkevich1980}. While this effect does not preclude neutrino
dark matter, it implies a top-down scenario for structure formation
where large structures form first, later fragmenting into smaller
ones. It was soon realized that the predicted properties of the
large-scale matter distribution did not seem to agree with
observations and that neutrino dark matter was apparently ruled out
\citep{White1983}.

Today it is widely accepted that the universe has critical density and
that its matter inventory sports several nontrivial
components. Besides some 5\% baryonic matter (most of it dark) there
are some 25\% cold dark matter in an unidentified physical form and
some 70\% of a negative-pressure component (``dark energy'').  And
because neutrinos do have mass, they contribute at least 0.1\% of the
critical density. This fraction is based on a hierarchical mass
scenario with $m_3=50~{\rm meV}$, the smallest value consistent with
atmospheric neutrino oscillations.

\begin{figure}[b]
\begin{center}
\epsfig{file=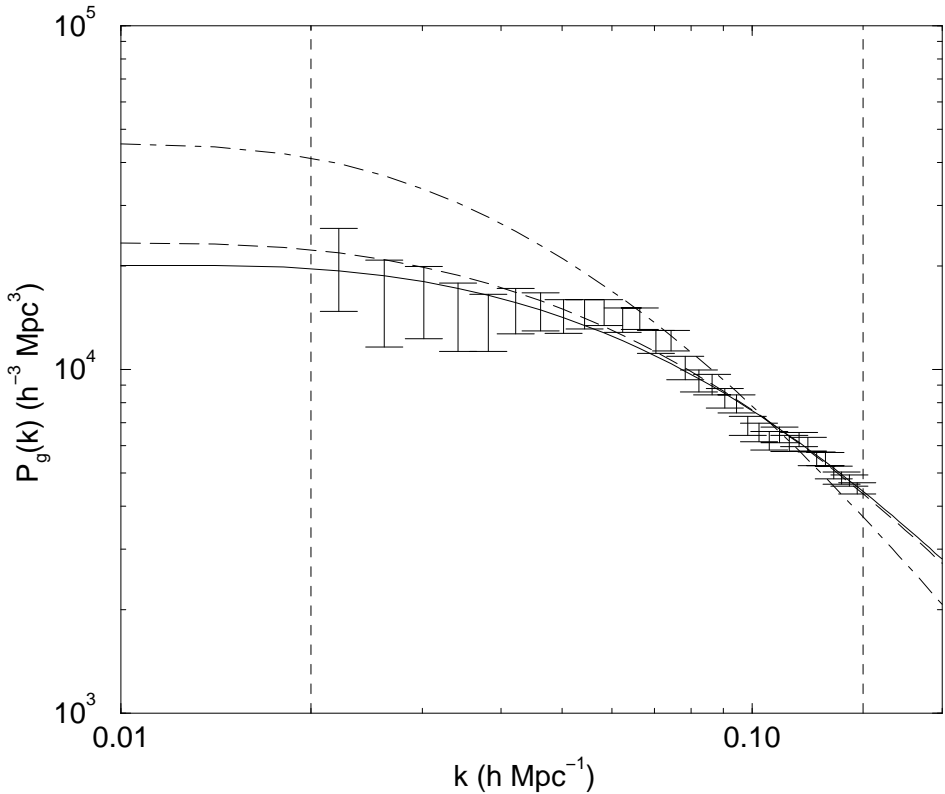,width=0.9\textwidth}
\end{center}
\caption{Power spectrum of the galaxy distribution function measured
by the 2dF Galaxy Redshift Survey. The solid line is the theoretical
prediction without neutrino dark matter ($\Omega_\nu=0$), the dashed
line for $\Omega_\nu=0.01$, and dot-dashed for $\Omega_\nu=0.05$. The
other cosmological parameters are $\Omega_{\rm M}=0.3$,
$\Omega_{\Lambda}=0.7$, $h=0.70$, and $\Omega_{\rm B}h^2=0.02$.
[Figure from \citet{Elgaroy2002} with permission.]}
\label{fig:1}
\end{figure}

An upper limit on the neutrino dark matter fraction can be derived
from the measured power spectrum $P_{\rm M}(k)$ of the cosmic matter
distribution. Neutrino free streaming suppresses the small-scale
structure by an approximate amount \citep{Hu1998}
\begin{equation}
\frac{\Delta P_{\rm M}}{P_{\rm M}}\approx 
-8\,\frac{\Omega_\nu}{\Omega_{\rm M}}
\end{equation}
where $\Omega_{\rm M}$ is the cosmic mass fraction in matter, i.e.\
excluding the dark energy. This effect is illustrated in
Fig.~\ref{fig:1} where $P_{\rm M}(k)$ measured by the 2dF Galaxy
Redshift Survey is shown and compared with the predictions for a cold
dark matter cosmology with neutrino fractions $\Omega_\nu=0$, 0.01,
and 0.05, respectively \citep{Elgaroy2002}.  When the theoretical
curves are normalized to the power at large scales, neutrinos indeed
suppress $P_{\rm M}(k)$ at large $k$.

Based on the 2dFGRS data, \citet{Elgaroy2002} and
\citet{Hannestad2002} find limits on $\sum m_\nu$ in the range
1.8--3.0~eV, depending on the assumed priors for other cosmological
parameters, notably the Hubble constant, the overall matter fraction
$\Omega_{\rm M}$, and the tilt of the spectrum of primordial density
fluctuations. For a reasonable set of priors one may adopt
\begin{equation}\label{eq:cosmicmasslimit}
\sum_{\rm flavors} m_\nu < 2.5~{\rm eV}
\end{equation}
at a statistical confidence level of 95\%. This limit corresponds
approximately to the dot-dashed ($\Omega_\nu=0.05$) curve in
Fig.~\ref{fig:1}, i.e.\ neutrinos may still contribute as much as 5\%
of the critical density, about as much as baryons.

Within the framework of the standard theory of structure formation,
the largest systematic uncertainty comes from the unknown biasing
parameter $b$ which relates the power spectrum of the galaxy
distribution to that of the true underlying matter distribution,
$P_{\rm Gal}(k)=b^2\,P_{\rm M}(k)$. The biasing parameter is one of
the quantities which must be taken into account when fitting all
large-scale structure data to observations of the galaxy distribution
and of the temperature fluctuations of the cosmic microwave background
radiation. In future the Sloan Digital Sky Survey will have greater
sensitivity to the overall shape of $P_{\rm Gal}(k)$ on the relevant
scales, allowing one to disentangle more reliably the impact of $b$
and $\Omega_\nu$ on $P_{\rm M}(k)$.  It is foreseen that one can then
reach a sensitivity of $\sum m_\nu \sim 0.65~{\rm eV}$ \citep{Hu1998}.

For a degenerate neutrino mass scenario the limit of
Eq.~(\ref{eq:cosmicmasslimit}) corresponds to a limit on the overall
mass scale of $m_\nu<0.8~{\rm eV}$, far more restrictive than the
laboratory limit Eq.~(\ref{eq:tritiumlimits}). However, the KATRIN
project for improving the tritium endpoint sensitivity is foreseen to
reach the 0.3~eV level \citep{Osipowicz2001}, similar to the
anticipated sensitivity of future cosmological observations.  If both
methods yield a positive signature, they will mutually re-enforce each
other. If they both find upper limits, again they will be able to
cross-check each other's constraints. 


\section{How Many Neutrinos in the Universe?}

The laboratory limits or future measurements of $m_\nu$ and the
cosmological limits or future discovery of a hot dark matter component
can be related to each other if the cosmic neutrino density $n_\nu$ is
known. However, the cosmic neutrino background can not be measured
with foreseeable methods so that one depends on indirect arguments for
determining $n_\nu$. Even if we accept that there are exactly three
neutrino flavors as indicated by the $Z^0$ decay width and that they
were once in thermal equilibrium does not fix $n_\nu$. Each flavor is
characterized by an unknown chemical potential $\mu_\nu$ or a
degeneracy parameter $\xi_\nu=\mu_\nu/T$, the latter being a quantity
invariant under cosmic expansion. While the observed baryon-to-photon
ratio suggests that the degeneracy parameters of all fermions are very
small, for neutrinos this is an assumption and not an established
fact.

In the presence of a degeneracy parameter $\xi_\nu$ the number and
energy densities of relativistic neutrinos plus anti-neutrinos in
thermal equilibrium are
\begin{eqnarray}
n_\nu&=&T_\nu^3\,\frac{3\zeta_3}{2\pi^2}\,
\left[1+\frac{2\ln(2)\,\xi_\nu^2}{3\zeta_3}
+\frac{\xi_\nu^4}{72\,\zeta_3}+O(\xi_\nu^6)\right]\,,\\
\rho_\nu&=&T_\nu^4\,\frac{7\pi^2}{120}\,
\left[1+\frac{30}{7}\left(\frac{\xi_\nu}{\pi}\right)^2
+\frac{15}{7}\left(\frac{\xi_\nu}{\pi}\right)^4\right]\,.
\end{eqnarray}
Therefore, if chemical potentials are taken to be the only uncertainty
of the cosmic neutrino density, $n_\nu$ can only be larger than the
standard value. In this sense the structure formation limits on the
hot dark matter fraction provide a conservative limit on the neutrino
mass scale $m_\nu$.  Conversely, a laboratory limit on $m_\nu$ does
not limit the hot dark matter fraction while a positive future
laboratory measurement of $m_\nu$ provides only a lower limit on
$\Omega_\nu$.

Big-bang nucleosynthesis (BBN) is affected by $\rho_\nu$ in that a
larger neutrino density increases the primordial expansion rate,
thereby increasing the neutron-to-proton freeze-out ratio $n/p$ and
thus the cosmic helium abundance. Therefore, the observed helium
abundance provides a limit on $\rho_\nu$ which corresponds to some
fraction of an effective extra neutrino species. In addition, however,
an electron neutrino chemical potential modifies
$n/p\propto\exp(-\xi_{\nu_e})$. Depending on the sign of $\xi_{\nu_e}$
this effect can increase or decrease the helium abundance and can
compensate for the $\rho_\nu$ effect of other flavors
\citep{Kang1992}. 
If $\xi_{\nu_e}$ is the only chemical potential, BBN provides
the limit
\begin{equation}
-0.01<\xi_{\nu_e}<0.07.
\end{equation}
Including the compensation effect, the only upper limit on the
radiation density comes from precision measurements of the power
spectrum of the temperature fluctuations of the cosmic microwave
background radiation and from large-scale structure measurements. A
recent analysis yields the allowed regions \citep{Hansen2002}
\begin{equation}
-0.01<\xi_{\nu_e}<0.22\,,\qquad |\xi_{\nu_{\mu,\tau}}|<2.6\,,
\end{equation}
in agreement with similar results of \citet{Hannestad2001} and
\citet{Kneller2001}.

However, the observed neutrino oscillations imply that the individual
flavor lepton numbers are not conserved and that in true thermal
equilibrium all neutrinos are characterized by one single chemical
potential $\xi_\nu$. If flavor equilibrium is achieved before $n/p$
freeze-out the restrictive BBN limit on $\xi_{\nu_e}$ applies to all
flavors, i.e.\ $|\xi_\nu|<0.07$, implying that the cosmic number
density of neutrinos is fixed to within about 1\%. In that case the
relation between $\Omega_\nu$ and $m_\nu$ is uniquely given by the
standard expression Eq.~(\ref{eq:Oega_nu}).

\begin{figure}[b]
\begin{center}
\epsfig{file=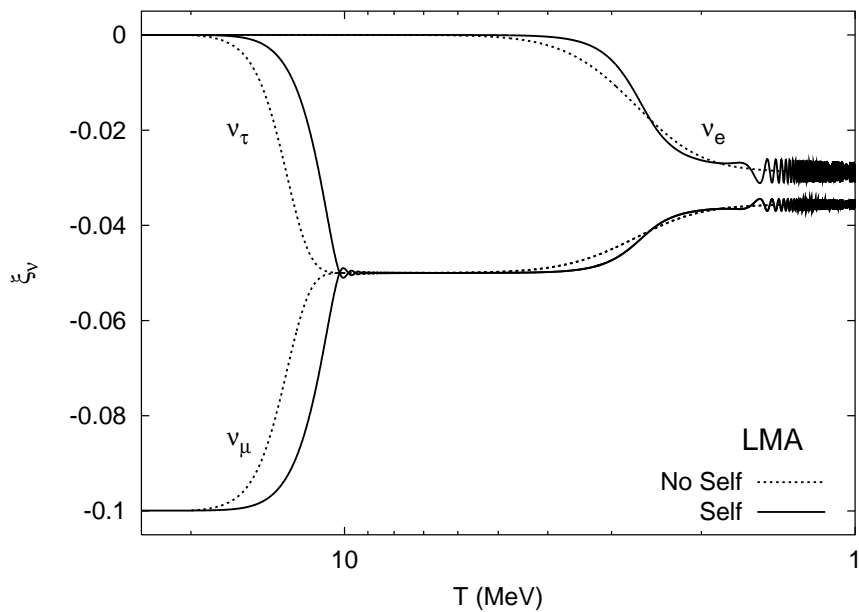,width=0.82\textwidth}
\end{center}
\caption{Cosmological evolution of neutrino degeneracy parameters
assuming the initial values $\xi_{\nu_e}=\xi_{\nu_\tau}=0$ and a
non-zero value for $\xi_{\nu_\mu}$.  The neutrino mixing parameters
were chosen according to the atmospheric and solar LMA solutions, and
taking $\Theta_{13}=0$.  [Figure from \citet{Dolgov2002} with
permission.]}
\label{fig:2}
\end{figure}

The approach to flavor equilibrium in the early universe by neutrino
oscillations and collisions was recently studied by
\citet{Lunardini2001}, \citet{Dolgov2002}, \citet{Wong2002}, and
\citet{Abazajian2002}. Assuming the atmospheric and solar LMA
solutions, an example for the cosmic flavor evolution is shown in
Fig.~\ref{fig:2}. The detailed treatment is rather complicated and
involves a number of subtleties related to the large weak potential
caused by the neutrinos themselves as they oscillate. The intriguing
phenomenon of synchronized flavor oscillations
\citep{Samuel1993,Pastor2002} plays an important and subtle role.

The practical bottom line, however, is rather simple. Effective flavor
equilibrium before $n/p$ freeze-out is reliably achieved if the solar
oscillation parameters are in the favored LMA region. In the LOW
region, the result depends sensitively on the value of the small but
unknown third mixing angle $\Theta_{13}$. In the SMA and VAC regions,
which are now heavily disfavored, equilibrium is not
achieved. Therefore, establishing LMA as the correct solution of the
solar neutrino problem amounts in our context to counting the number
of cosmological neutrinos and thus to establishing a unique
relationship between the neutrino mass scale $m_\nu$ and the cosmic
neutrino density $\Omega_\nu$. A final confirmation of LMA is expected
by the Kamland reactor experiment \citep{Shirai2002} within the next
few months of this writing.


\section{{\boldmath$Z$}-Burst Scenario for the Highest-Energy 
Cosmic Rays}

The number density and mass of cosmic background neutrinos is also
relevant for the propagation of extremely high-energy (EHE) neutrinos
that may be produced by hitherto unknown astrophysical sources at
cosmological distances. Assuming that neutrinos with energies in the
$10^{21}$--$10^{22}$~eV range are somewhere injected in the universe,
and assuming that the cosmic background neutrinos have masses in the
neighborhood of 1~eV, the center-of-momentum energy is in the
neighborhood of the $Z^0$ boson mass. Put another way, the required
neutrino energy for the $Z^0$-resonance is
\begin{equation}
E_\nu=\frac{m_Z^2}{2m_\nu}=
\frac{4.2\times10^{21}~{\rm eV}}{m_\nu/{\rm eV}}\,.
\end{equation}
The subsequent decay of the $Z^0$-bosons on average produces two
nucleons, 10~$\pi^0$ mesons which subsequently decay into photons, and
17~$\pi^\pm$ mesons which subsequently decay into $e^\pm$ and
neutrinos. These $Z$-bursts would be a source for cosmic rays at the
upper end of the observed spectrum, i.e.\ with energies at and above
$10^{20}~\rm eV$ \citep{Weiler1999,Fargion1999,Fodor2002}.

The motivation for considering this sort of scenario is the difficulty
of explaining the observed highest-energy cosmic rays which exceed the
Greisen-Zatsepin-Kuzmin (GZK) cutoff, i.e.\ which can not reach us
from large distances because the cosmic microwave background renders
the universe opaque for protons around and above $10^{20}~\rm eV$. No
compelling explanation for the super-GZK cosmic rays exists, leaving
much room and motivation for speculations and exotic scenarios
\citep{Sigl2001}. Of course, the $Z$-burst scenario is not a real
explanation because it appeals to unknown EHE neutrino sources
producing huge fluxes and with rather exotic properties
\citep{Kalashev2002}.

\begin{figure}[t]
\begin{center}
\epsfig{file=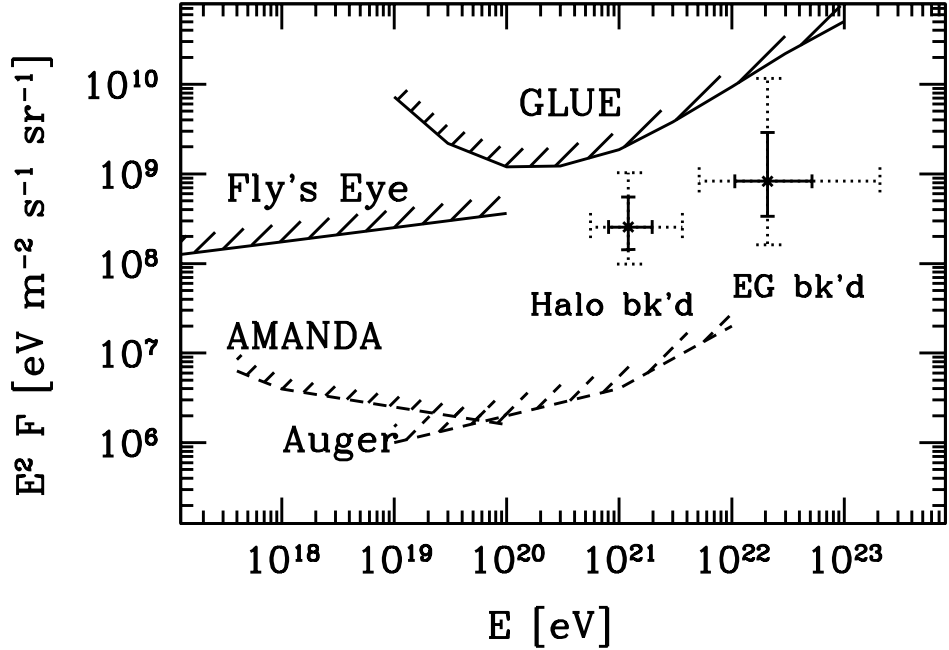,width=0.9\textwidth}
\end{center}
\caption{Extremely high-energy neutrino fluxes. The crosses represent
two scenarios for explaining the highest-energy cosmic rays by
$Z$-bursts. The upper hatched curves are the current experimental
limits, the lower ones the foreseen sensitivity of future experiments.
[Figure from \citet{Fodor2002} with permission.]}
\label{fig:3}
\bigskip
\end{figure}

However, one should view this scenario as an opportunity. The required
EHE neutrino fluxes are below current limits, but are accessible to
future experiments such as the IceCube neutrino telescope at the South
Pole or the Auger air shower array in Argentina (Fig.~\ref{fig:3}).
Therefore, if such fluxes are detected they provide a handle on sub-eV
masses of the cosmic background neutrinos by virtue of the $Z$-burst
production of super-GZK cosmic rays. This would be the only evidence
for the cosmic neutrino background and its properties other than
provided by big-bang nucleosynthesis.

\section{Leptogenesis}

Neutrino masses in the sub-eV range can play an interesting albeit
indirect role for creating the baryon asymmetry of the universe (BAU)
in the framework of leptogenesis scenarios \citep{Fukugita1986}.  The
main ingredients are those of the usual see-saw scenario for small
neutrino masses. Restricting ourselves to a single family, the
relevant parameters are the heavy Majorana mass $M$ of the ordinary
neutrino's right-handed partner and a Yukawa coupling $g_\nu$ between
the neutrinos and the Higgs field $\Phi$.  The observed neutrino then
has a Majorana mass
\begin{equation}
m_\nu=\frac{g_\nu^2\langle\Phi\rangle^2}{M}
\end{equation}
that can be very small if $M$ is large, even if the Yukawa coupling
$g_\nu$ is comparable to that for other fermions.  Here,
$\langle\Phi\rangle$ is the vacuum expectation value of the Higgs
field which also gives masses to the other fermions.

The heavy Majorana neutrinos will be in thermal equilibrium in the
early universe. When the temperature falls below their mass, their
density is Boltzmann suppressed. However, if at that time they are no
longer in thermal equilibrium, their abundance will exceed the
equilibrium distribution. The subsequent out-of-equilibrium decays can
lead to the net generation of lepton number. CP-violating decays are
possible by the usual interference of tree-level with one-loop
diagrams with suitably adjusted phases of the various couplings. The
generated lepton number excess will be re-processed by standard-model
sphaleron effects which respect $B-L$ but violate $B+L$. It is
straightforward to generate the observed BAU by this mechanism.

The requirement that the heavy Majorana neutrinos freeze out before
they get Boltzmann suppressed implies an upper limit on the
combination of parameters $g_\nu^2/M$ that also appears in the see-saw
formula for $m_\nu$. The out-of-equilibrium condition thus implies an
upper limit on $m_\nu$.  Detailed scenarios for generic neutrino mass
and mixing schemes have been worked out, see \citet{Buchmuller2000}
for a recent review and citations of the large body of pertinent
literature.

The bottom line is that neutrino mass and mixing schemes suggested by
the atmospheric and solar oscillation data are nicely consistent with
plausible leptogenesis scenarios. Of course, it is an open question of
how one would go about to verify or falsify leptogenesis as the
correct baryogenesis scenario. Still, it is intriguing that massive
neutrinos may have a lot more to do with the baryons than with the
dark matter of the universe!


\section{Time-of-Flight Dispersion of Supernova Neutrinos}

In principle, neutrino masses can be measured by the dispersion of a
neutrino burst from a pulsed astrophysical source, notably a supernova
(SN). The time-of-flight delay of massive neutrinos with energy
$E_\nu$ is
\begin{equation}
\Delta t = \frac{m_\nu^2}{2E_\nu^2}\,D
\end{equation}
where $D$ is the distance to the source. Therefore, if a neutrino
burst has the intrinsic duration $\Delta t$ and the energies are
broadly distributed around some typical energy $E_\nu$, one is
approximately sensitive to masses
\begin{equation}
m_\nu > 10~{\rm eV}\,\left(\frac{E_\nu}{10~\rm MeV}\right)\,
\left(\frac{\Delta t}{\rm s}\right)^{1/2}
\left(\frac{10~\rm kpc}{D}\right)^{1/2}\,.
\end{equation}
The measured $\bar\nu_e$ burst of SN~1987A had 
$E\approx 20~{\rm MeV}$, $\Delta t\approx 10~{\rm s}$, and $D\approx
50~{\rm kpc}$, leading to the well-known limit $m_{\nu}\alt20~{\rm
eV}$ \citep{Loredo1989}.  In a recent re-analysis \citet{Loredo2002}
find a somewhat more restrictive limit. Either way, these results are
only of historical interest because the tritium and cosmological
limits are now much more restrictive.

The neutrino burst from a future galactic SN could yield more
restrictive limits because one would expect up to 8000 events in a
detector like Super-Kamiokande for a typical galactic distance of
around 10~kpc.  With such a high-statistics signal the relevant
time-scale $\Delta t$ is the fast rise-time of around 100~ms rather
than the overall burst duration of several seconds.  Therefore, one is
sensitive to smaller masses than the SN~1987A burst, despite the
shorter baseline. From detailed Monte-Carlo simulations
\citet{Totani1998} infers that Super-Kamiokande would be sensitive to
about $m_{\nu}\agt3~{\rm eV}$, almost independently of the exact
distance.  (At a larger distance one gains baseline but loses
statistics, two effects that cancel for a given detector size.)

Conceivably this sensitivity could be improved if a gravitational wave
signal could be detected preceding the neutrinos, signifying the
instant of the stellar collapse \citep{Fargion1981,Arnaud2002}. In
this case one may be sensitive to about 1~eV.

It is also conceivable that a SN collapses to a black hole some short
time after the original collapse. In this case the neutrino signal
would abruptly terminate (within $\Delta t\alt0.5~{\rm ms}$), thereby
defining a very short time scale.  \citet{Beacom2000,Beacom2001}
found that Super-Kamiokande would be sensitive to $m_{\nu}\agt1.8~{\rm
eV}$.

In the foreseeable future megatonne neutrino detectors may be
constructed to search for proton decay and to perform precision
long-baseline oscillation measurements with neutrino beams.  Such
detectors would have about 30 times the fiducial volume of
Super-Kamiokande.  The exact $m_\nu$ sensitivity for such an
instrument has not been worked out.  With a megatonne detector one
could measure SN neutrinos throughout the local group of
galaxies. From Andromeda at a distance of 750~kpc one would get around
50 events.  Using the overall signal duration for $\Delta t$ yields a
sensitivity in the few eV range.

The only conceivable time-of-flight technique that could probe the
sub-eV range involves Gamma-Ray Bursts (GRBs) which have been
speculated to be strong neutrino sources. If the neutrino emission
shows time structure on the millisecond scale, and assuming a
cosmological distance of 1~Gpc, one would be sensitive to neutrino
masses $m\agt 0.1~{\rm eV}~E/{\rm GeV}$. Therefore, observing
millisecond time structure in sub-GeV neutrinos from a GRB would be
sensitive to the sub-eV mass scale
\citep{Halzen1996,Choubey2002}.

\newpage


\section{Discussion and Summary}

The compelling detection of flavor oscillations in the solar and
atmospheric neutrino data have triggered a new era in neutrino
physics. In the laboratory one will proceed with precision experiments
aimed at measuring the details of the mixing matrix. Future tritium
decay experiments may well be able to probe the overall neutrino mass
scale down to the 0.3~eV range, but if the absolute masses are
smaller, it will be very difficult to measure them, and the overall
mass scale may remain the most important unknown quantity in neutrino
physics for a long time to come.

Unfortunately, it is unlikely that astrophysical time-of-flight
methods will help much. Foreseeable SN neutrino detectors are
sensitive to eV masses, but not the sub-eV range. On the bright side
this means that the measured neutrino light-curve of a future galactic
SN will faithfully represent the source without much modifications by
neutrino dispersion.

Cosmological large-scale structure data at present provide the most
restrictive limit on neutrino masses of $\sum m_\nu<2.5~{\rm eV}$,
corresponding to $m_\nu<0.8~{\rm eV}$ in a degenerate mass scenario. A
rigorous relationship between the cosmic hot dark matter fraction
$\Omega_\nu$ and $m_\nu$ depends on the cosmic neutrino
density~$n_\nu$. If the solar LMA solution is correct, big-bang
nucleosynthesis constrains $n_\nu$ without further assumptions about
the neutrino chemical potentials. In the LMA case neutrinos reach
de-facto flavor equilibrium before the epoch of weak-interaction
freeze out.

While neutrinos do not play a dominant role for dark matter or
structure formation, the mass and mixing schemes suggested by the
oscillation experiments are nicely consistent with leptogenesis
scenarios for creating the cosmic baryon asymmetry. Therefore, massive
neutrinos may be closely related to the baryons in the universe, not
the dark matter.

If extremely-high energy neutrinos will be observed in future, the
$Z$-burst scenario provides a handle on the cosmic background
neutrinos and their mass through the observed cosmic rays near the GZK
cutoff.

The great advance in our knowledge of neutrino properties together
with the cosmological precision information has rendered the question
of neutrino masses in astrophysics and cosmology far more subtle than
it looked only a few years ago.  We will not know if the accumulation
of new results would have persuaded Dennis Sciama that nature has used
massive neutrinos perhaps in different ways for cosmology than he
himself had imagined for so long.  Either way, the connection between
neutrino properties and astroparticle physics remains of fundamental
interest.

\newpage


\section*{Acknowledgments}

This work was supported, in part, by the Deutsche Forschungsgemeinschaft
under grant No.\ SFB-375 and by the European Science Foundation (ESF)
under the Network Grant No.~86 Neutrino Astrophysics.



\end{document}